\documentclass[a4paper,11pt]{article}
\pdfoutput=1 

\usepackage{jcappub} 

\usepackage[T1]{fontenc} 

\usepackage{multirow}
\usepackage[english]{babel}
\usepackage{amsmath}
\usepackage{amssymb}
\usepackage{epsfig}
\usepackage{graphics,psfrag,rotating}
\usepackage{graphicx}
\usepackage{dcolumn}
\usepackage{bm}
\usepackage{slashbox}
\usepackage{array,multirow}
\usepackage{xspace}
\usepackage{color}

\newcommand{\lsim}   {\mathrel{\mathop{\kern 0pt \rlap
  {\raise.2ex\hbox{$<$}}}
  \lower.9ex\hbox{\kern-.190em $\sim$}}}
\newcommand{\gsim}   {\mathrel{\mathop{\kern 0pt \rlap
  {\raise.2ex\hbox{$>$}}}
  \lower.9ex\hbox{\kern-.190em $\sim$}}}

\newcommand{\Ep}{E_{\bar p}}
\title{\boldmath Antiproton signatures from astrophysical and dark matter sources at the galactic center}

\author[a,b,c]{J. A. R. Cembranos}
\author[a]{ V. Gammaldi}
\author[a]{A.\,L.\,Maroto}

\affiliation[a]{Departamento de F\'{\i}sica Te\'orica I, Universidad Complutense de Madrid, E-28040 Madrid, Spain;}
\affiliation[b]{Facultad de Ciencias, CUICBAS, Universidad de Colima, 28040 Colima, Mexico;}
\affiliation[c]{Dual C-P Institute of High Energy Physics, 28040 Colima, Mexico.}

\emailAdd{cembra@ucm.es}
\emailAdd{vivigamm@ucm.es}
\emailAdd{maroto@fis.ucm.es}

\abstract{The center of our Galaxy is a complex region characterized by extreme phenomena. The presence of the supermassive Sagittarius A* black hole, a high dark matter density and an even higher baryonic density are able to produce very energetic processes. Indeed, high energetic gamma-rays have been observed by different telescopes, although their origin is not clear. In this work, we estimate the possible antiproton flux component associated with this signal. The expected secondary astrophysical antiproton background already saturates the observed data. It implies that any other important astrophysical source leads to an inconsistent excess.
We estimate the sensitivity of PAMELA to this new primary antiproton source, which depends on the diffusion model and its spectral features. In particular, we consider antiproton spectra described by a power-law, a monochromatic signal and a Standard Model particle-antiparticle  channel production. This latter spectrum is typical in the production from annihilating or decaying dark matter. We pay particular attention to the case of a heavy dark matter candidate, which could be associated with the High Energy Stereoscopic System (HESS) data observed from the J1745-290 source.}

\begin{document}
\maketitle
\flushbottom

\section{Introduction}

The Galactic Center (GC) hosts large macroscopic concentrations of gas, Dark Matter (DM) and interstellar radiation, which
implies an important diffuse Galactic emission in this region. In addition, the GC contains a large number of resolved and
unresolved sources of cosmic-rays. Such a complex structure copiously sources different cosmic-rays from hadronic inelastic interactions, charged particle acceleration,
inverse Compton scattering and bremsstrahlung.  This dense environment does not allow to reconstruct cosmic-ray fluxes
from first principles without non-trivial extrapolations and important assumptions.

Different studies of the GC region have found interesting features in the spectra of cosmic-ray fluxes,
mainly related to gamma-ray emissions, and reported as excesses with respect to expected backgrounds.
Some of them, as the one observed by the EGRET telescope in the diffuse gamma-ray emission  \cite{Hunger:1997we, deBoer:2005tm}, has
been fully explained  as having a systematic origin~\cite{Stecker:2007xp} since it has not been confirmed by other data \cite{Abdo:2010nz}
as the one collected by the Fermi Large Area Telescope (LAT) \cite{Vitale, ferm}.
However, it has been claimed that these new data contain another possible excess \cite{Hooper:2010mq}.
At higher energies, observations of gamma-rays from the GC have been reported by several collaborations such as CANGAROO \cite{CANG}, VERITAS \cite{VER}, HESS  \cite{Aha, HESS}, MAGIC \cite{MAG}.  Also neutrino fluxes are expected to be originated from the GC. In fact,
the IceCube collaboration have reported the observation of 37 high-energy neutrinos. They seem to have an astrophysical origin and 5 of them are likely originated from the GC \cite{neutrino}.

There are different potential candidates for the primary source of new cosmic-rays over the aforementioned backgrounds. In particular, there
are point sources associated with the Sgr A* black hole \cite{Boyarsky:2010dr, Chernyakova:2011zz},
supernova remnants such as the Sgr A East supernova \cite{SgrA}, unresolved populations of millisecond pulsars \cite{Abazajian:2010zy}, or other unidentified point sources. As we have mentioned, the majority of the observations are related to gamma-ray fluxes, but it is expected that the same primary source, which constitutes the origin of the observed signal, may produce leptonic or hadronic counterparts. In addition, the production of a concrete particle will induce the production of others depending on the particular type of particle and energy. This secondary production would affect mainly the diffuse signal through hadronic emission by inelastic proton collision with the interstellar gas, inverse Compton scattering of interstellar radiation by cosmic-ray electrons and positrons, or bremsstrahlung \cite{Stecker:1977ua, Abdo:2010nz}.

All these effects make the analysis very challenging. In order to model the GC background, different templates have been used. However, significant systematic effects are associated with cosmic-ray density distribution and diffuse hadronic emission~\cite{Linden:2010ea}. Indeed, there is a large ignorance about cosmic-rays in this region since different populations of cosmic-rays are likely to inhabit the GC itself and may have an important contribution to the fluxes observed at the Earth. Another fundamental issue with using the diffusion models is the set of templates employed to reproduce the morphology of the hadronic and inverse Compton Galactic emission. For example, gas column-density map templates neglect the possibility of an enhanced cosmic-ray abundance in the inner Galaxy, and the inverse-Compton template depends strongly on specific choices for the input parameters in the {\tt Galprop} code \cite{Strong:1998fr}.

With all the commented caveats in mind, we will assume that such emission from the GC exists and we will estimate the possibility of detecting such signal under the assumption that it is very localized around the GC. In particular, we will focus on an antiproton emission.
The $e^\pm$ and $p\bar p$ data from ATIC/PPB-BETS, PAMELA, FERMI and AMS have been largely studied. It has been speculated during the
last years about the possibility of explaining the leptonic data at high energy with DM annihilation or decay.
However, the data are also consistent with astrophysical primary sources \cite{AMS-astro}. On the contrary, antiproton observations seem perfectly consistent with astrophysical expectations, whose origin is due to the interactions between cosmic-ray nuclei and the Interstellar Matter (ISM). In this sense, antiproton data can be used to characterize diffusion models of charged particles along our galaxy, or to constrain new physics, whose antiproton flux may be identified up the diffusion background. This is the case of DM models, whose indirect astrophysical searches are fundamental in order to investigate the constraints and the prospectives for the detection of different DM models \cite{DMcr, HDMW, antipJ, cosmics,antiprotonsCR}. This is particularly true for heavy DM candidates, whose observation in direct detection experiments or particle accelerators \cite{lab} is highly challenging.

In concrete, we will use the PAMELA antiproton data \cite{PAM}, that are in perfect agreement with \textit{secondary} and \textit{tertiary} antiprotons production. Astrophysical uncertainties due to the antiproton diffusion model affect the antiproton flux at the Top Of the Atmosphere (TOA) and several standard and non-standard diffusion models can be found in the literature \cite{antipJ, HE}. In this work, we will study the propagation of the antiproton flux emitted by a localized source at Galactic Center (GC) and the prospective antiproton flux for different emission spectra. Our aim is to study the prospective signature that could arise in the antiproton flux due to such a source. Restrictions can be set depending on the total integrated flux and the features of the emission spectra. The manuscript is organized as follow: In Section II, we will review the antiproton diffusion equation and its solution for the particular case of a point-like source at the GC. In Section III, we will analyze the prospective flux at TOA produced by a fiducial power law and monochromatic antiproton spectra for such a point-like source at the GC.
Section IV will be devote to the study of a heavy DM candidate able to explain the gamma-ray emission from the same region and
detected by HESS \cite{HESSfit}. Finally, we will summarize our main results in Section V.

\section{Antiproton propagation}
Charged cosmic-ray propagation in the Galaxy is a complex process affected by many different physical phenomena. Propagation parameters are set by B/C and sub-Fe/Fe cosmic-ray nuclei data analyses. Different configuration of parameters may be compatible with both set of data \cite{HE}. Antiproton energy losses, convention and reacceleration also affect the flux at the TOA. Energy losses are mainly due to two effects: First, ionization in neutral ISM and ionized plasma; and second, the existence of a Galactic wind. The latter phenomena is described as a constant convective wind velocity $V_c$, that pushes the antiprotons far away from the Galactic plane. In the middle of this plane, at z=0, a singularity takes place since $V_c$ has opposite sign above and below the Galactic plane \cite{HE}. The Galactic wind is due to
a constant flow of irregularities in the Galactic magnetic field and it cannot be neglected
in the central part of the Milky Way. This fact can be deduced from observations from ROSAT
and Fermi  \cite{Crocker:2010qn,Lacki:2013zsa}. However, there is not a concrete implementation
of this effect for the antiproton diffusion model within this region, and we will not take it
into account in our analysis. We have computed explicitly the consequences of an important
increase for the convective velocity in the model and it produces a significant loss
of antiprotons, mainly, in the low part of the spectrum. The systematic errors introduced
in this way are comparable to the existing uncertainties in the diffusion model itself.

The interaction between charged particles and inhomogeneities is described by the pure space
diffusion coefficient $K(E)$. This term is energy-dependent because higher energy particles
are sensitive to larger spatial scales:
\begin{equation}
K(\Ep)=K_0\beta(p/\text{GeV})^\Delta.
\end{equation}
Here, $p=(\Ep^2+2m_pE_{\bar p})^{1/2}$ is the antiproton ($\bar p$) momentum; $\beta=v_{\bar p}/c=(1-m_p^2/(\Ep+m_p)^2)^{1/2}$, its velocity ($c=3\times10^5$ km/s ); $\Ep=E-m_p$, the $\bar p$ kinetic energy; and $m_p=0.938$ GeV/$c^2$, its mass ($c=1$ in atomic units). The parameters $K_0$ and $\Delta$ depend on the diffusion model, and parameterize the antiproton escape probability from the confinement volume. This volume is identified with the Galactic halo, that is described as a cylinder of radius $R_D$, and halo half-hight $L$. The Galactic plane at $z=0$ can be modeled as a thin disk of thickness $2\,h=200\,\text{pc}$. The antiproton number at density per unit energy $f(t,\vec r, \Ep )=dN_{\bar p}/d\Ep$ vanishes on the surface of the cylinder at height $z=\pm L$, and at radius $r=R_D$.  The larger is $L$ and $R_D$, the larger the probability for particles emitted in remote sources to reach us \cite{Donato, Cirelli}. 
In Tab. \ref{K(E)}, we show the parameters for models with minimum, medium and maximum
propagation consistent with the commented observations \cite{Cirelli, K(E)}. We will use these models
for our analyses although they are optimized for cosmic-ray species produced following the
distribution of supernova remnants in the galaxy. The extrapolation of such models to study
the GC region is another source of systematic uncertainties. Indeed, diffusion effects for
antiprotons at distances around 1 pc from the GC may be negligible since they can lose their
energy {\it in situ} by synchrotron radiation due to the very large value of the turbulent magnetic
field within this region \cite{Melia}. One should take into account that this hypothesis
could suppress the sensitivity to antiproton fluxes originated at the GC by several orders of
magnitude \cite{Bringmann:2014lpa}.
  
\begin{table}[tb]
\centering
\begin{tabular}{|c|c|c|c|c|}
\hline
\hline
Model & $\Delta$ & $K_0\left[\text{kpc}^2/\text{Myr}\right]$ & $V_c\left[\text{km/s}\right]$&L $\left[\text{kpc}\right]$\\
\hline
\hline
MIN  & 0.85 & 0.0016 & 13.5 & 1\\
\hline
MED  & 0.70 & 0.0112 & 12 & 4\\
\hline
MAX & 0.46 & 0.0765 & 5 & 15 \\
\hline
\end{tabular}
\caption{Value of the parameters associated with different diffusion models of the antiproton propagation within the Milky Way.
The adjectives minimum, medium and maximum refer to the probability of such diffusion \cite{K(E)}. }
\label{K(E)}
\end{table}

Standard sources and interactions with the ISM are confined to the thin disk. The $p\bar p$ interaction in the Galactic plane depends on the inelastic and spallation cross section. We will use \cite{Cirelli}:
\begin{equation}
\sigma_{p\bar p}^\text{inel}\left(\Ep\right)=24.7\left[1+0.584(\Ep/\text{GeV})^{-0.115}+0.856(\Ep/\text{GeV})^{-0.566}\right]\text{mbar}\,.
 \end{equation}
Such an expression for the inelastic annihilation cross section $\sigma_{p\bar p}^\text{inel}=\sigma_\text{ann}+\sigma_\text{not-ann}$
is an extrapolation of the behaviour at low energies ($\Ep\lesssim 100$ GeV) that is consistent with data \cite{HE} in order to describe the interstellar $p\bar p$ interactions in the galactic plane, both as proton-antiproton annihilation and proton-hydrogen scattering (\textit{secondary} contribution) or energy loss (\textit{tertiary} contribution). Within this approximation, we are neglecting the \textit{tertiary} antiprotons that lose a significant fraction of their energy. At high energies the $\sigma_\text{ann}$ tends to be smaller, so that
$\sigma_{p\bar p}^\text{inel}\simeq\sigma_\text{not-ann}$. In any case, the exact expression does not affect the final result at these
higher energies in an appreciable way. This is because the \textit{tertiary} contribution does not produce new antiprotons,
but merely redistributes them towards lower energies \cite{Cirelli, HE}.
All these contributions and the \textit{primary} source $\hat Q(t, \vec r, \Ep)$ are transported according to the diffusion equation:
\begin{eqnarray}\frac{\partial}{\partial t}\frac{dN_{\bar p}}{d E_{\bar p}}-K(E_{\bar p})
\cdot\bigtriangledown^2\frac{dN_{\bar p}}{d E_{\bar p}} +\frac{\partial}{\partial z}\left(sign(z)\frac{dN_{\bar p}}{d E_{\bar p}}V_{c}\right)
=\hat Q(t,\vec r, \Ep)-2h\delta(z)\Gamma_\text{inel}\left(\Ep\right)\frac{dN_{\bar p}}{d E_{\bar p}}\,.
\label{eq:eqf}
\end{eqnarray}
We will analyze steady states defined by $\partial f/\partial t=0$ and $\hat Q(t,\vec r, \Ep)=\hat Q(\vec r, \Ep)$.
In addition, we will assume that the \textit{primary} source can be factorized in two functions depending on
its spatial distribution ($Q_X$) and its spectral shape ($Q_E$) in the following way: $\hat Q(\vec r, \Ep)= Q_X(\vec r)\cdot Q_E(\Ep)$.
In such a case, a general solution of Eq. (\ref{eq:eqf})
for the antiproton flux per unit of energy and per steradian can be written as:
\begin{equation}
\frac{d\Phi_{\bar p}^{\text{Sun}}}{d\Ep}=\frac{v_{\bar p}}{4\pi} \frac{dN_{\bar p}^{\text{Sun}}}{d\Ep}
= \frac{v_{\bar p}}{4\pi}R(r_{\odot},\Ep)Q_E(\Ep)\,,
\label{eq:flux}
\end{equation}
where $dN^{\text{Sun}}_{\bar p}/d\Ep$ is the solution at $r=r_\odot$ and
\begin{equation}
R(r_\odot, \Ep)= \sum_{m=1}^\infty J_0\left(\zeta_m\frac{r_\odot}{R_D}\right)\text{exp}\Big[-\frac{V_cL}{2K\left(\Ep\right)}\Big]
\frac{\Pi_m\left(L,\Ep\right)}{A_m\left(\Ep\right)\text{sinh}\left[S_m\left(\Ep\right)L/2\right]}\,
\label{eq:R}
\end{equation}
describes the Galactic antiproton production and propagation with
\begin{eqnarray}
A_m\left(\Ep\right)&=&2h\Gamma_\text{inel}\left(\Ep\right)+V_c+K\left(\Ep\right)S_m\left(\Ep\right)\text{coth}\big[S_m\left(\Ep\right)L/2\big]\,,\\
S_m\left(\Ep\right)&=&\left(\frac{V_c^2}{K\left(\Ep\right)^2}+\frac{4\zeta_m^2}{R_D^2}\right)^{1/2}\,.
\end{eqnarray}
Because of the cylindrical symmetry, solutions are found in terms of Bessel functions
of order n-th ($J_n$) and thus the properties of these Bessel functions
control the behavior of these steady-state solutions. In particular, $J_0$ is the zero-$th$ order Bessel function and $\zeta_m$ is its $m$-$th$ order zero. On the other hand, $\Pi_m$ depends also on the Bessel function of first order $J_1$:
 \begin{eqnarray}
 \Pi_m\left(L,\Ep\right)=\frac{2}{J_1^2\left(\zeta_m\right)R_D^2}\int^R_0dr\,rJ_0\left(\zeta_m\frac{r}{R_D}\right)\times\,\,\,\,\,\,\,\,\,\,\,\,\,\,\,\,\,\,\,\, \\ \nonumber
 \,\,\,\,\,\,\,\int^L_{-L}dz\,\text{exp}\Big[\frac{V_c(L-z)}{2K\left(\Ep\right)}\Big]\text{sinh}\left[S_m\left(\Ep\right)(L-z)/2\right]Q_X(\vec r)\,.
 \label{eq:P}
 \end{eqnarray}
As we commented in the introduction, we would like to estimate constraints and prospectives for the detection of antiproton fluxes originated
from the inner part of our Galaxy. In order to simplify the mathematical treatment, we will assume a localized source at the GC. In particular, we will model $Q_X(\vec r)$  with a point-like spatial distribution described by a three dimensional $\delta$-function ($\delta^{(3)}$)
centered at GC:
\begin{equation}
Q_X(\vec r)=Q_X^0 \delta^{(3)}(\vec r)\equiv Q_X^0  \frac{1}{2\pi r}\delta^{(1)}(r)\delta^{(1)}(z)\,,
\end{equation}
where $Q_X^0$ is a normalization constant, and we have explicitly written $\delta^{(3)}(\vec r)$ in terms of one dimensional
$\delta$-functions ($\delta^{(1)}$) in cylindrical coordinates $(r,\theta,z)$. In such a case, Equations (\ref{eq:R}) and (\ref{eq:P})
acquire simpler expressions:
\begin{eqnarray}
 \Pi_m^{\delta}\left(L,\Ep\right)&=&\frac{Q_X^0}{{\pi} R_D^2}\frac{J_0(0)}{J_1^2\left(\zeta_m\right)}\text{exp}\Big[\frac{V_c\,L}{2K\left(\Ep\right)}\Big]
 \text{sinh}\left(S_m\left(\Ep\right) L/2\right)\,,\\
 R^{\delta}(r_\odot, \Ep)&=&\frac{Q_X^0}{\pi R_D^2}\sum_{m=1}^\infty\frac{J_0\left(\zeta_{m}\frac{r_\odot}{R_D}\right)}{A_m(\Ep)J_1^2\left(\zeta_m\right)}\,.
 \label{Rdelta}
 \end{eqnarray}
Therefore, the steady flux of antiprotons for a localized source located at the center of our galaxy can be written as:
\begin{equation}
\frac{d\Phi_{\bar p}^\delta}{d\Ep}=\frac{v_{\bar p}}{4\pi}R^\delta(r_\odot, \Ep)Q_E(\Ep),
\label{eq:fluxdelta}
\end{equation}
where $Q_E(\Ep)$ is its characteristic energy spectrum. \\

Equations (\ref{eq:flux}) and (\ref{eq:fluxdelta}) provide the solution at the position of the Sun of the antiproton diffusion equation in the Galaxy.
To get the antiproton flux at the TOA  we have taken into account the solar modulation by assuming the so-called force-field or Fisk potential \cite{Cirelli, Perko}:

\begin{equation}
\frac{d\Phi_{\bar{p}}^{\text{TOA}}}{d\Ep^{\text{TOA}}}=\frac{p^2_{\text{TOA}}}{p^2}\frac{d\Phi_{\bar p}}{d\Ep}
\end{equation}
with $\Ep=\Ep^{\text{TOA}}+|Ze|\phi_F$.
As it is well known, the particular value of the Fisk potential $\phi_F$ in order to parameterize the solar modulation on cosmic-rays
depends on the solar activity and the epoch of observation. We have used $\phi_F=0.5$ GV since
different works (for instance, read \cite{Cirelli:2014lwa}) have concluded that the range between 0.1 GV and 1.0 GV is the most appropriate
for the PAMELA data taking period. This effect is important at low energies. It reduces
the flux of antiprotons for energies below 10 GeV, and consequently, the sensitivity of the antiproton study if the
analysis is typically dominated by low-energy observations.

\section{Energy spectra associated with general astrophysical sources}

\begin{figure}
\begin{center}
 \begin{minipage}{.4\textwidth}
  \includegraphics[width=\textwidth,height=\textheight,keepaspectratio]{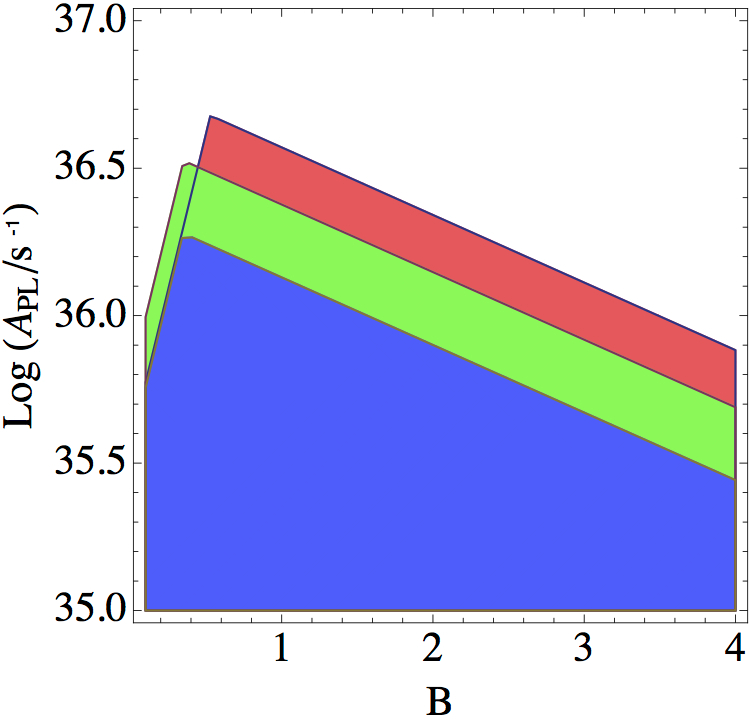}
\caption {\footnotesize{Sensitivity regions derived from PAMELA data for a point-like source of antiprotons at the GC characterized by a  power law
emission spectrum with total amplitude $A_{PL}$ and spectral index $-(B+1)$. Red, green and
blue regions correspond to  high, medium and low diffusion models respectively.
The experiment is sensitive in the upper region independently on the
diffusion features of the antiprotons within the Milky Way.
Low spectral indices are constrained by high energy data whereas the sensitivity
related to high spectral indices is determined by low energy data. It explains the
change in the slope shown by the Figure.}}
\label{plplot}
 \end{minipage}
 \hspace{2cm}
 \begin{minipage}{.4\textwidth}
  \includegraphics[width=\textwidth,height=\textheight,keepaspectratio]{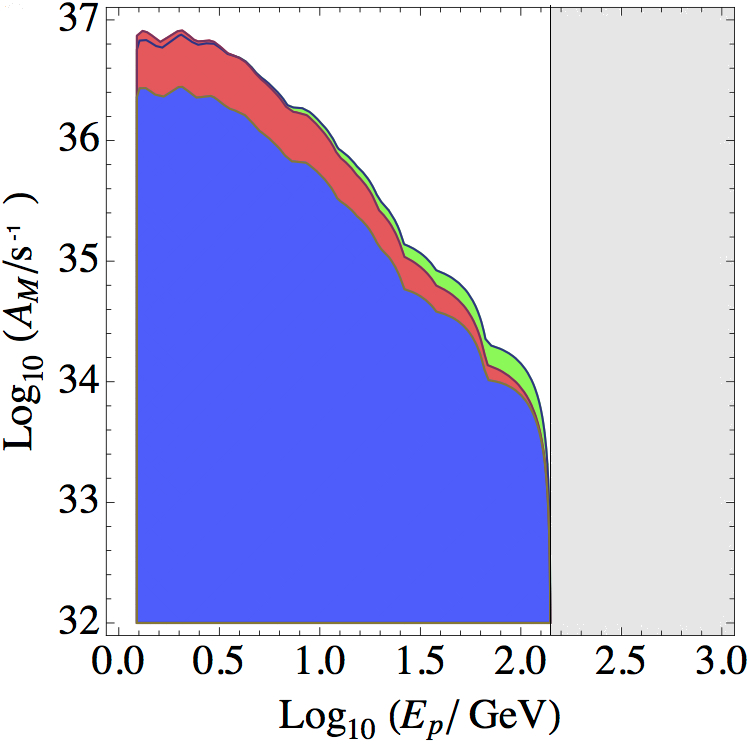}
\caption {\footnotesize{The same as in Fig. \ref{plplot} but for a point-like source with monochromatic emission, $\Ep^0$. The figure shows the sensitivity region of the parameter space $(A_M,\Ep^0)$ for all possible characteristic energies between $1$ GeV and $130$ GeV approximately.
Higher emission energies are not constrained by the lack of observational data.}}
\label{mono}
 \end{minipage}
 \end{center}
\end{figure}

As we have commented in the introduction, the GC hosts different types of point-like sources such as black holes, supernovas, pulsars, etc.
These sources have been identified mainly by observations of their gamma-ray emissions. In order to estimate the possible observation
of their antiproton counterpart, it is necessary to assume a particular spectral shape without entering into the details of the particular
source. If the range of antiproton energies, that is relevant for the analysis, does not extend for many orders of magnitude, a power law
spectrum is typically a good approximation. Indeed, acceleration of cosmic-rays by Supernovae Remnats (SNR) or Pulsar Wind Nebulae (PWN) are examples of power law spectra (with a cut-off at higher energies \cite{AMS-astro}). In such a case, the antiproton source spectrum can be described
in terms of two parameters:
\begin{equation}
Q_E^{\text{PL}}(\Ep)=\frac{Q_E^{0-\text{PL}}}{\Ep}\cdot\left(\frac{\Ep}{\text{GeV}}\right)^{-B}\,,
\label{pl}
\end{equation}
where $B$ characterizes the suppression of power at high energies, and $Q_E^0$ normalizes the spectral flux. Therefore, we can write the
total antiproton flux at TOA in terms of two constants:
\begin{equation}
\frac{d\Phi_{\bar p}^{\delta-\text{PL}}}{d\Ep}=\frac{v_{\bar p}}{4\pi}R^\delta(\Ep)\frac{A_\text{PL}}{\Ep}\cdot\left(\frac{\Ep}{GeV}\right)^{-B}\,,
\label{plTOA}
\end{equation}
where $A_{\text{PL}}\equiv Q_X^0\cdot Q_E^{0-\text{PL}}$, takes into account the total normalization of the emission. We have compared such flux with the antiproton data observed by PAMELA. In order to be conservative, we will neglect the background contribution and assume that observations are sensitive to the signal if it produces a higher antiproton flux that the observed one at any point of the spectrum.
We have checked that a complete study with the use of a realistic background as the one given in \cite{Kappl:2014hha} gives very similar results.
By following this approach, the sensitivity on the amplitude and spectral index are shown in Fig. \ref{plplot} by using different diffusion models. In any case, such a dependence is not significant as it can be seen in the mentioned figure.
\begin{figure}[t]
\begin{center}
\epsfxsize=13cm
\resizebox
{10cm}{7.5cm}
{\includegraphics{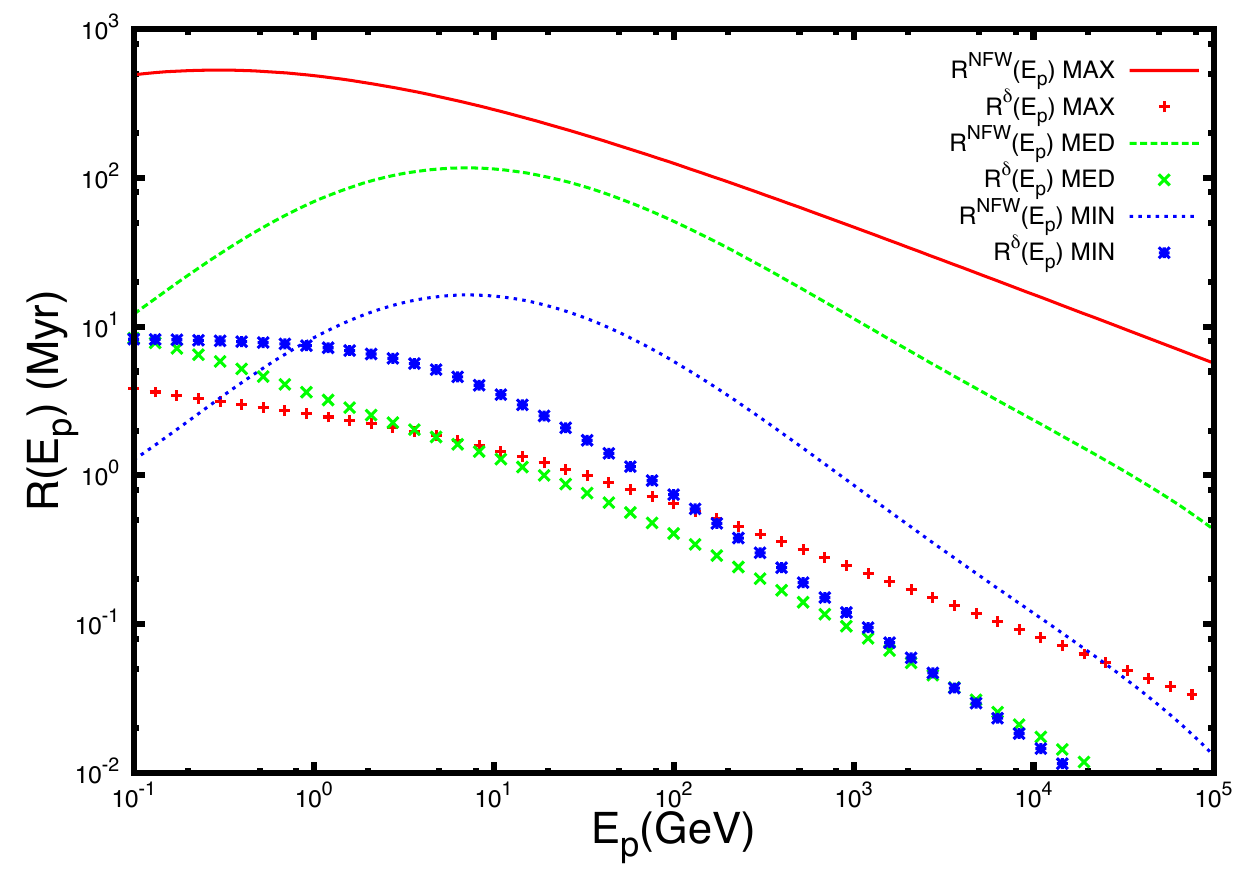}}
\caption {\footnotesize{Dependence of the antiproton diffusion function $R(\Ep)$, in terms of the antiproton kinetic energy $\Ep$,
for the maximum, medium and minimum diffusion models, associated with the annihilating DM distributed by following a NFW profile
($R^\text{NFW}(\Ep)$ represented by a full, broken and dotted line respectively) as in \cite{Cirelli}; and a point-like DM distribution at the GC ($R^\delta(\Ep)$ represented by plus, cross and star points respectively, with $Q_X^{0-\text{NFW}}=2.13\cdot10^{60}\,\text{m}^3$ sr as in \cite{HESSfit}.}}
\label{Rfunction}
\end{center}
\end{figure}

The power law spectral source is constrained fundamentally by the low energy data ($\Ep\lesssim 1$ GeV). However, in a general case, observations
of antiprotons at higher energies can be more relevant. We can analyze this effect by assuming a monochromatic source. In this sense, this spectrum
characterizes complementary features to the power law assumption. In particular, we will describe the spectral shape as a gaussian distribution in energy
\begin{equation}
Q_E^{\text{M}}(\Ep)=\frac{Q_E^{{0-\text{M}}}}{\sqrt{2\pi}\Delta_E}e^{-\frac{(\Ep-\Ep^0)^2}{2\,\Delta_E^2}}\,,
\end{equation}
with the standard deviation given by the typical energy resolution of the device. For the PAMELA calorimeter the energy resolution is of the order of $5\%$ of the antiproton energy ($\Delta_E \simeq 0.05 \Ep$) for a large spectral range. In this case, the two parameters that define the spectrum are the spectral normalization $Q_E^0$ and the monochromatic emission energy $\Ep^0$. We can perform an analogous analysis that for the power law
shape. In this case, the total antiproton signal at TOA reads:
\begin{equation}
\frac{d\Phi_{\bar p}^{\delta-\text{M}}}{d\Ep}=\frac{v_{\bar p}}{4\pi}R^\delta(\Ep)
\frac{A_{{\text{M}}}}{\sqrt{2\pi}\Delta_E}e^{-\frac{(\Ep-\Ep^0)^2}{2\,\Delta_E^2}}\,,
\label{plTOA}
\end{equation}
where the two constants, that parameterize the signal are $A_{{\text{M}}}\equiv Q_X^0\cdot Q_E^{{0-\text{M}}}$,
and $\Ep^0$.
Again, the analysis shows very low dependence with the diffusion model (see Fig. \ref{mono}).
However, in this case, all the observational data are important depending on $\Ep^0$,
and in fact, data at high energies are most constraining since the observed antiproton flux is much more reduced.
For energies higher than $\Ep^0\simeq 130$ GeV (up the black solid vertical line), mono-energetic sources are unconstrained
by PAMELA observations due to the lack of data.

\begin{figure}
\begin{center}
 \begin{minipage}{.4\textwidth}
  \includegraphics[width=\textwidth,height=\textheight,keepaspectratio]{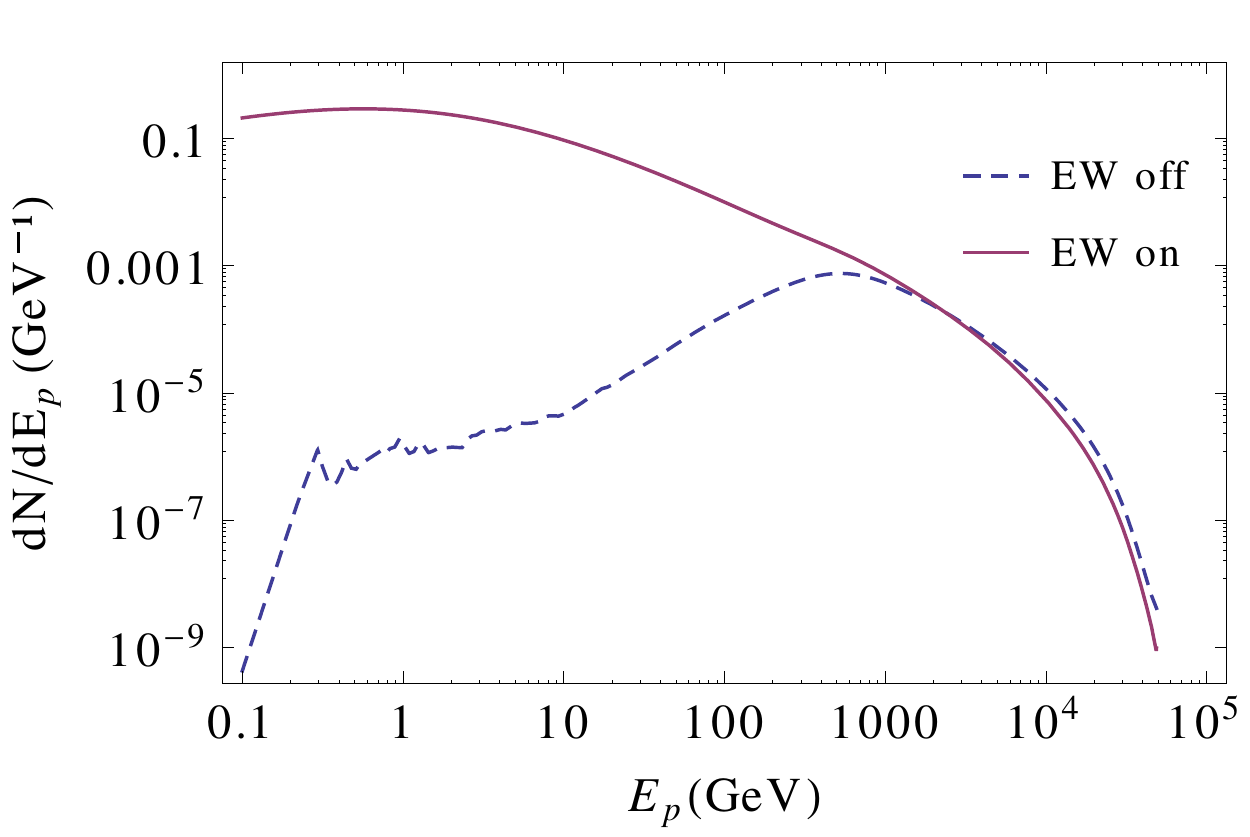}
\caption {\footnotesize{Antiproton differential flux at the source for DM annihilating into $W^+W^-$ channel
with mass $m_{\text{DM}}=48.8$ TeV, before propagation. It is evident that the electroweak (EW) radiation effects
cannot be neglected.}}
\label{Wpsource}
 \end{minipage}
 \hspace{2cm}
 \begin{minipage}{.4\textwidth}
  \includegraphics[width=\textwidth,height=\textheight,keepaspectratio]{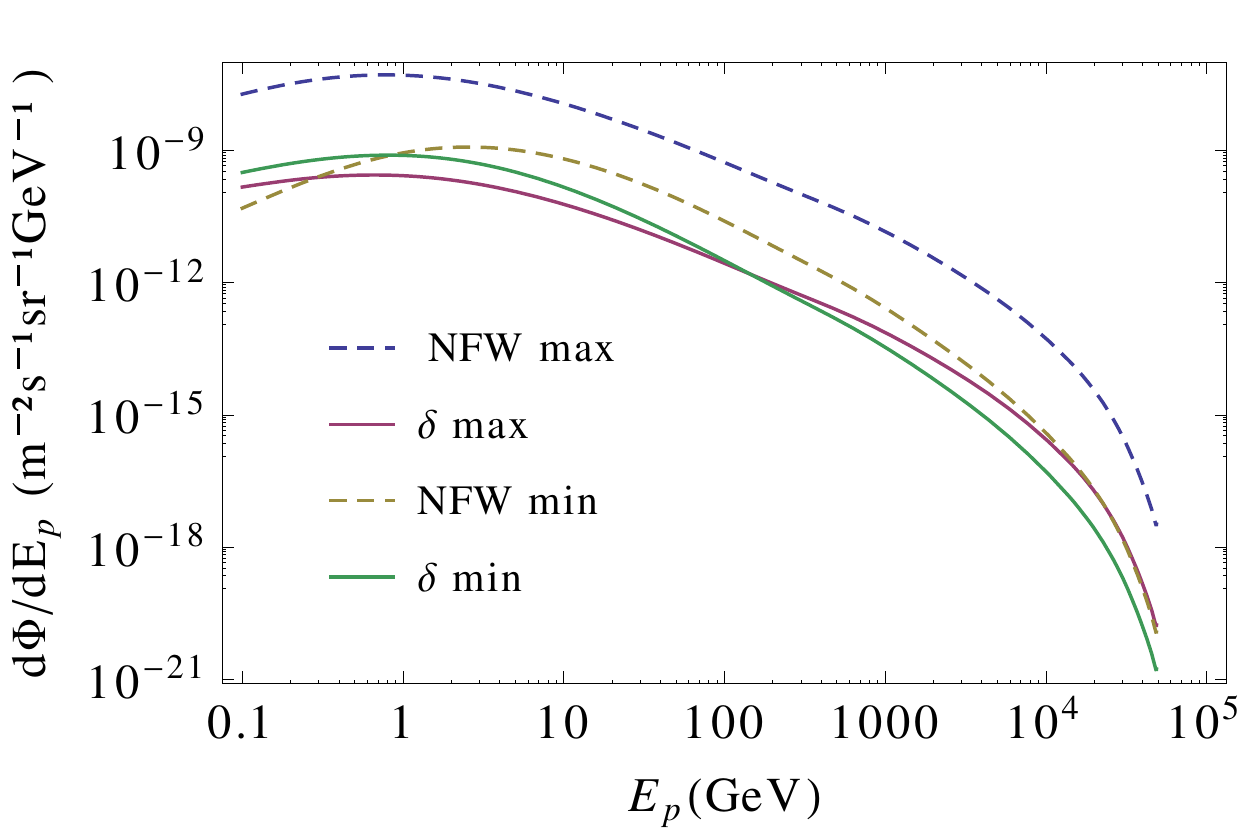}
\caption {\footnotesize{Antiproton differential flux at the TOA after propagation for
$48.8$ TeV DM annihilating into $W^+W^-$ pairs.
Dashed lines correspond to a spatial distribution following the standard NFW halo profile:  upper and lower lines
stand for the maximum and minimum propagation model, respectively.
The full lines mean the same for antiproton propagation from a point-like source located at the GC with amplitude $Q_X^{0-\text{NFW}}=2.13\cdot10^{60}\,\text{m}^3$ sr.}}
\label{TOAflux}
 \end{minipage}
 \end{center}
\end{figure}

\section{Dark Matter and the HESS gamma-ray J1745-290 source}

Another interesting spectral shape to be studied is the one associated with annihilation or decay of DM particles, which can cluster around a
very compact region of the center of our galaxy. For example, baryonic effects may modify the gravitational potential by increasing the
density in the GC \cite{Blumenthal,Prada:2004pi} and compressing the dark halo. This scenario is under debate \cite{Romano}, but it could
enhance the importance of the GC region for indirect DM searches, and in particular, for the antiproton analysis.
In fact, it has been shown in \cite{HESSfit} that the J1745-290 HESS gamma-ray data \cite{Aha, HESS} are well fitted as a point-like DM source at the GC. This analysis shows good agreement with DM annihilation or decay into $u\bar u$, $d\bar d$, $s\bar s$ and $t\bar t$ quark-antiquark channels and $W^+W^-$ and $ZZ$ boson channels. Leptonic and other quark-antiquark channels were excluded with $95.4\%$ confidence level. Therefore, a significant flux of antiprotons is expected to be produced if the DM is the origin of this gamma-ray emission.

On the other hand, there is another motivation to consider this type of localized DM sources at the GC. Antiproton fluxes from DM have been largely studied in previous works \cite{DMcr, HDMW, morselli}, but they have been focused on the total dark halo contribution, which is dominated
by local density contributions. Indeed, numerical results for general antiproton fluxes at the TOA generated by annihilation or
decay of DM particles in the Galaxy halo with different propagation models have been provided in \cite{Cirelli}. It has been shown
that numerical computations of different solutions of the diffusion equation present an unavoidable singularity around the GC, because of the
central steepness, which usually characterizes DM halo density profiles. In particular, in Ref. \cite{Cirelli}, this divergence
is replaced by a well behaved approximation below an arbitrary critical radius of few parsecs (footnote 16 in \cite{Cirelli}).
In this sense, the addition of a DM contribution from a point-like source at the GC
provides a more complete analysis (for a different way of regularizing the central DM halo singularity, read \cite{Bringmann:2006im}).

\begin{figure}
\begin{center}
 \begin{minipage}{.4\textwidth}
  \includegraphics[width=\textwidth,height=\textheight,keepaspectratio]{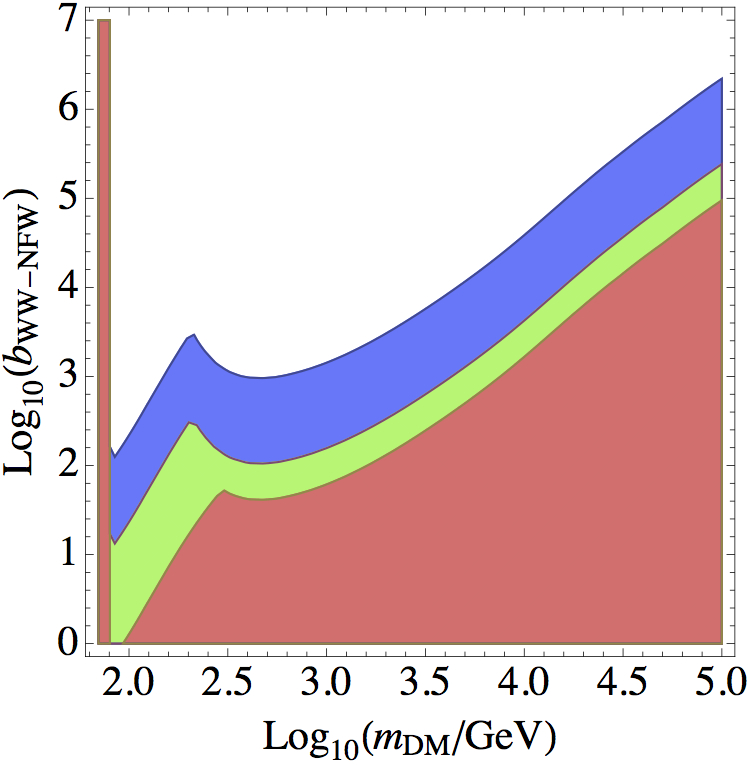}
\caption {\footnotesize{
Sensitivity regions from PAMELA data for DM annihilating into $W^+W^-$ pairs distributed in a NFW halo.
The experiment is sensitive to the upper (white) region independently on the
diffusion features of the antiprotons within the Milky Way.
On the contrary, the lower (red) region is allowed.
Intermediate regions (green and blue) are allowed also
for medium and minimum diffusion models, respectively
(the boost factor associated to the NFW halo emission is plotted in the vertical axis $b^{WW}_\text{NFW}$,
whereas the boost factor associated to a potential contribution from the GC is zero: $b^{WW}_{\delta-\text{NFW}}=0$).
Figure starts near the $W^+W^-$ direct production threshold (vertical line).
The y-axis can be understood as the DM cross-section divided by the standard thermal one used for reference
if annihilating DM distributed in a NFW profile with no substructure is assumed. However, the meaning
of the y-axis is more general since the boost factor can be associated
with a different dark halo, the presence of substructure or a possible
DM decay channel.
}}
\label{NFWDM}
 \end{minipage}
 \hspace{2cm}
 \begin{minipage}{.4\textwidth}
  \includegraphics[width=\textwidth,height=\textheight,keepaspectratio]{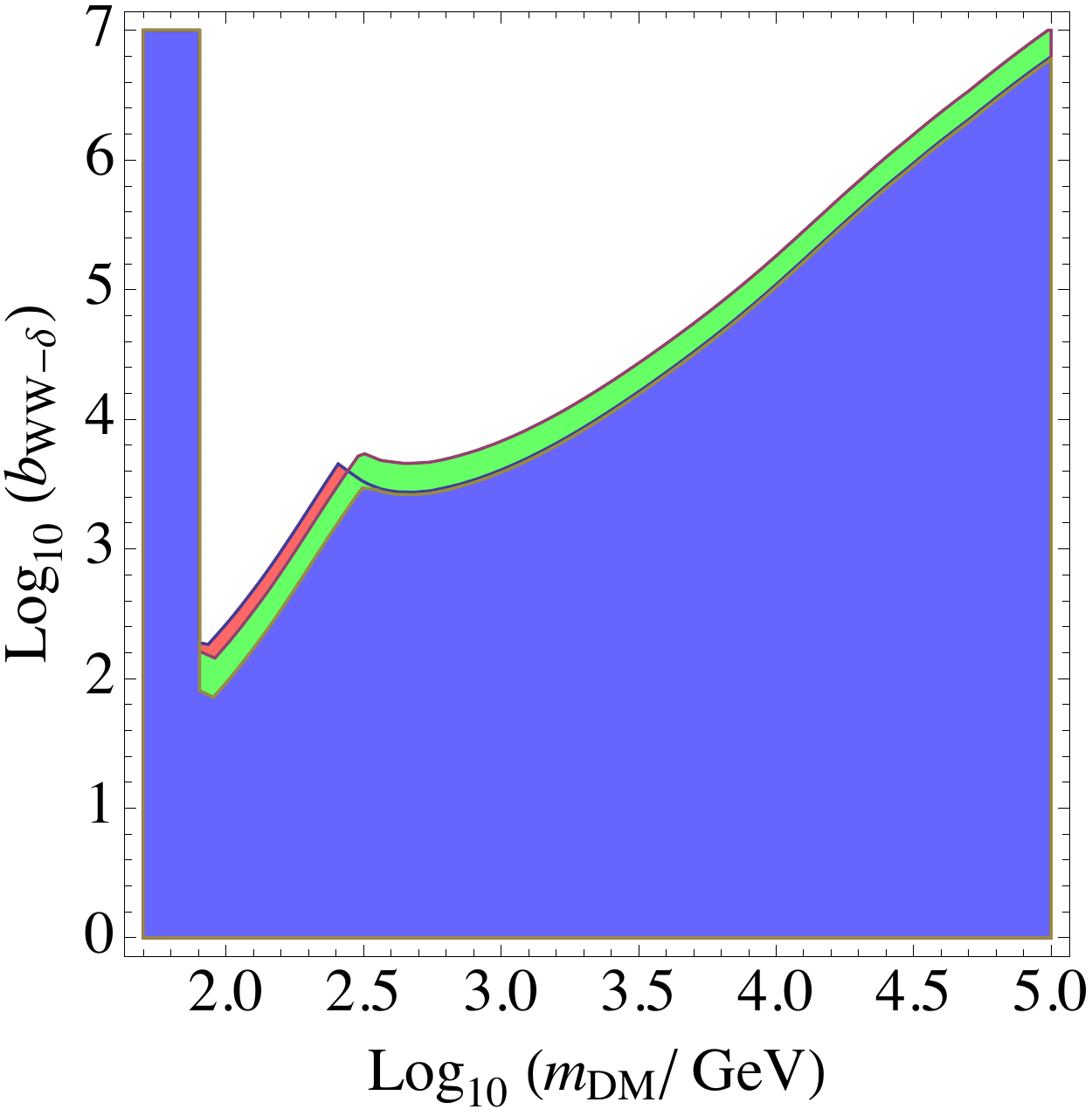}
\caption {\footnotesize{The same as in Fig. \ref{NFWDM} but for a point-like DM distribution at the GC
with amplitude $Q_X^0=b^{WW}_{\delta-\text{NFW}}\;Q_X^{0-\text{NFW}}$
(here, the boost factor associated to the NFW halo emission is taken to be zero: $b^{WW}_\text{NFW}=0$,
and the boost factor associated to the point-like source $b^{WW}_{\delta-\text{NFW}}$,
is plotted in the vertical axis).}}
\label{deltaDM}
 \end{minipage}
 \end{center}
 \end{figure}

 \begin{figure}
 \begin{center}
  \begin{minipage}{.4\textwidth}
  \includegraphics[width=\textwidth,height=\textheight,keepaspectratio]{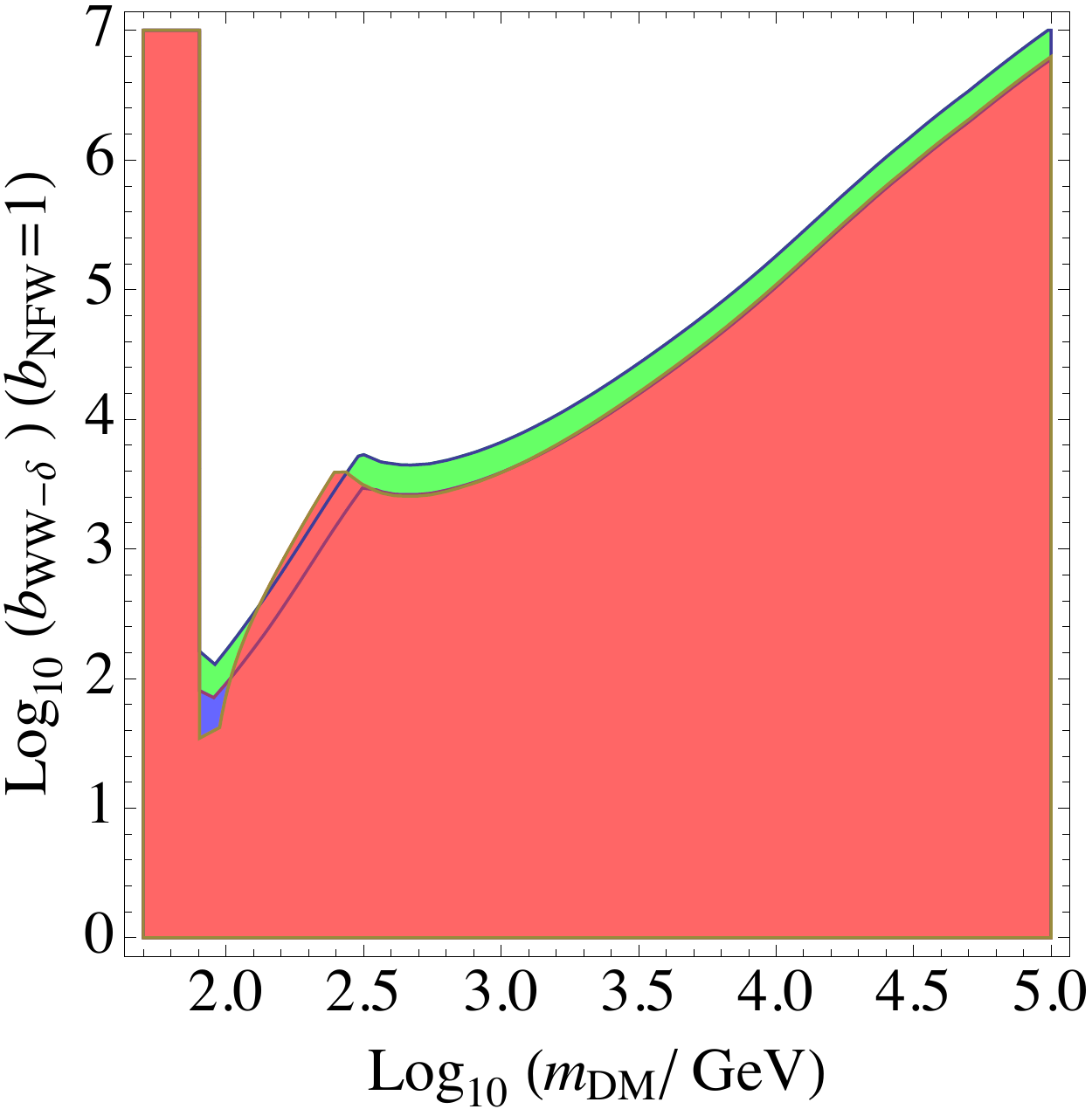}
\caption {\footnotesize{The same as in Fig. \ref{deltaDM} but with the addition of the contribution of the standard NFW halo.
(i.e., the boost factor associated with the NFW halo emission is taken to be one: $b^{WW}_\text{NFW}=1$,
and the boost factor associated with the point-like source $b^{WW}_{\delta-\text{NFW}}$,
is again plotted in the vertical axis).}}
\label{NFWdelta}
 \end{minipage}
  \hspace{2cm}
 \begin{minipage}{.4\textwidth}
  \includegraphics[width=\textwidth,height=\textheight,keepaspectratio]{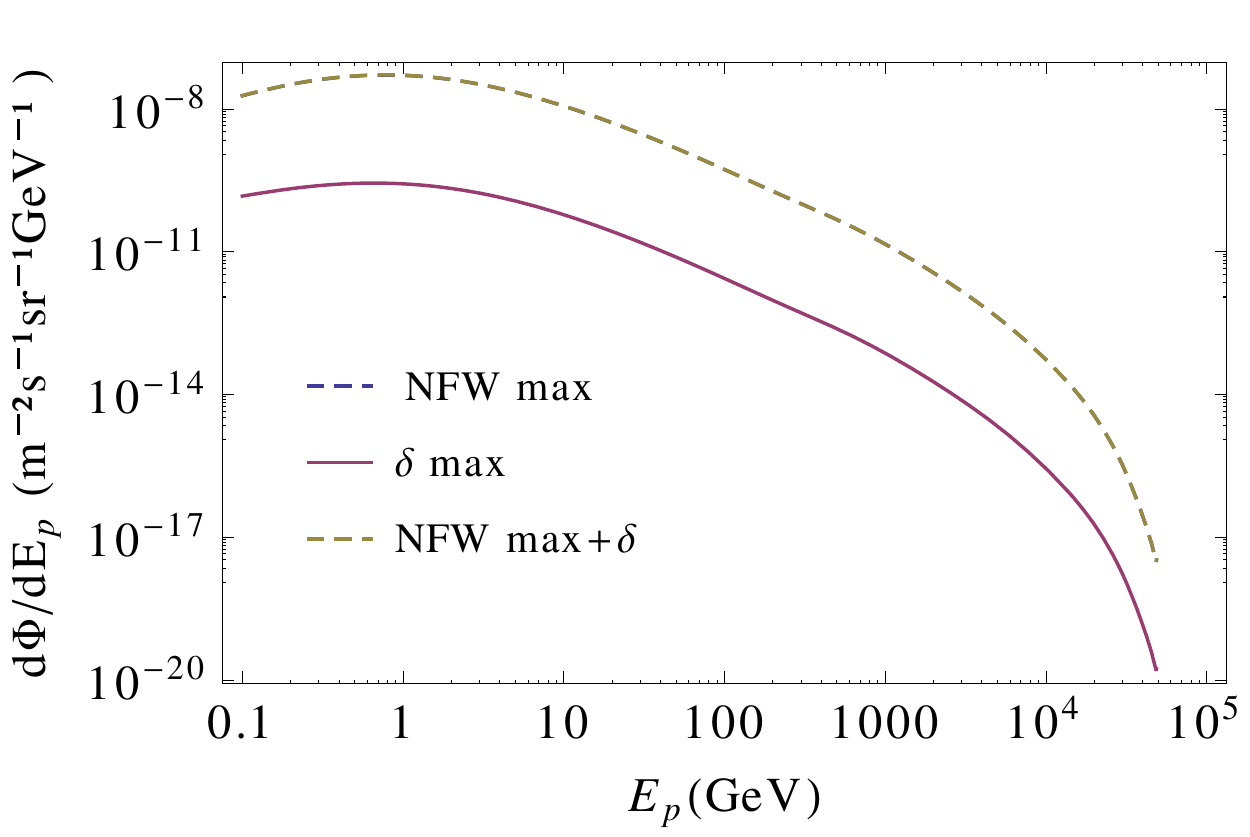}
\caption {\footnotesize{
Antiproton differential flux at the TOA after propagation for
$48.8$ TeV DM annihilating into $W^+W^-$ pairs by assuming the
maximum propagation model.
Dashed lines correspond to the contribution from a spatial distribution following the standard NFW halo profile,
whereas the full line stands for the contribution from a  point-like source located at the GC with amplitude $Q_X^{0-\text{NFW}}=2.13\cdot10^{60}\,\text{m}^3$ sr. This second contribution is subdominant and the addition
of both components overlap with the one coming from the continuous halo.}}
\label{NFWdeltamax}
 \end{minipage}
 \end{center}
\end{figure}

In order to give a well established physical reference for  the value of this point-like contribution, we will use the standard
contribution of annihilating DM from the NFW profile \cite{Navarro:1996gj} of our Galaxy and the gamma-ray observation in the direction of the GC.
In this case, the expression for the astrophysical factor is given by :
\begin{eqnarray}
\langle J \rangle= \frac{1}{\Delta\Omega}\int_{\Delta\Omega}\text{d}\Omega\int_0^{l_{max}(\Psi)} \rho^2 [r(l)] dl(\Psi)\,,
\label{astrofactor}
\end{eqnarray}
where $l$ is the distance from the Sun to a particular point of the DM halo, that is related to the radial distance $r$ from the GC,
through the equation: $r^2 = l^2 + D_\odot^2 -2D_\odot l \,\cos \Psi$, with $l_{max} = D_\odot \cos \Psi + \sqrt{r_{max}^2-D_\odot^2 \sin \Psi }$.  The distance between the solar system and the center of the Galaxy is $D_\odot\simeq 8.5$ kpc. We will assume the solid
angle $\Delta\Omega=10^{-5}$ sr.
In this case, the equivalent DM point-like source associated with the density distribution $\rho_\delta(\vec r)$ is a spatial $\delta$-function
centered at the GC, which in spherical coordinates $\vec r\equiv(r_s,\theta_s,\phi_s)$, can be written as:
\begin{equation}
Q_X(\vec r)=Q_X^0\delta^{(3)}(\vec r)\equiv Q_X^{0}  \frac{1}{4\pi r^2}\delta^{(1)}(r_s)
\equiv \left[\frac{\rho_\delta(\vec r)}{\rho\odot}\right]^2 \,,
\end{equation}
where we normalize this value to the local DM density $\rho_\odot\simeq 0.3\text{ GeVcm}^{-3}$. Thus, the
contribution of the point-like spatial distribution to the gamma-ray flux is
\begin{equation}
\langle J\rangle_{\Delta\Omega}=\frac{1}{\Delta\Omega}
\int_{\Delta\Omega} d\Omega\int_0^{l_{max}}\rho^2_\odot Q_X^0 \delta^{(3)}\big[\vec r(l)\big]dl(\psi)=\frac{Q_X^0}{\Delta\Omega}\left(\frac{\rho_\odot^2}{D_\odot^2}\right)\,.
\end{equation}

The same astrophysical factor for the gamma-ray observation coming from the GC direction with a telescope with the same HESS solid angle $\Delta\Omega\simeq 10^{-5}$ sr and NFW DM density distribution is
$\langle J\rangle^{\text NFW}_{\Delta\Omega^\text{HESS}}=280\times10^{23}\text{GeV}^2\text{cm}^{-5}$. Thus,
the equivalent normalization constant, that we will use as reference, is given by
\begin{eqnarray}
Q_X^{0-\text{NFW}}&=&\langle J\rangle^{NFW}_{\Delta\Omega}\Delta\Omega_\text{HESS}\left(\frac{D_\odot}{\rho_\odot}\right)^2
\simeq 2.13\cdot10^{60}\,m^3\,\text{sr}.
\end{eqnarray}

It is interesting to compare between the antiproton flux coming from the GC and the expected contribution from the continuous halo.
The relation will depend on the diffusion model and particular features of the DM density distribution. In order to give numerical
results, we will focus again in the annihilating DM case within the standard NFW profile. The diffusion Equation (\ref{eq:eqf})
applies to the total antiproton source coming from DM annihilation. The steady solution is also given by Eq. (\ref{eq:flux}), but
we need to take into account the continuous distribution of the DM to compute $R^\text{NFW}(\Ep)$ from Eq. (\ref{eq:R}):
\begin{equation}
Q_X(\vec r)= \left[\frac{\rho_\text{NFW}(\vec r)}{\rho\odot}\right]^2 \,.
\end{equation}
As we see in Fig. \ref{Rfunction}, when $Q_X^0\simeq 1$ in units of $Q_X^{0-\text{NFW}}$, the propagation function $R^\delta(\Ep)$ for the point-like DM source at the GC is comparable with the propagation function $R^\text{NFW}(\Ep)$ for the NFW halo profile. Moreover, fitting HESS data
requires typically $Q_X^0\simeq 10^3\,Q_X^{0-\text{NFW}}$, which implies that the GC contribution could dominate the standard one (see Fig.  \ref{Rfunction}).

\begin{figure}
 \begin{center}
 \begin{minipage}{.4\textwidth}
  \includegraphics[width=\textwidth,height=\textheight,keepaspectratio]{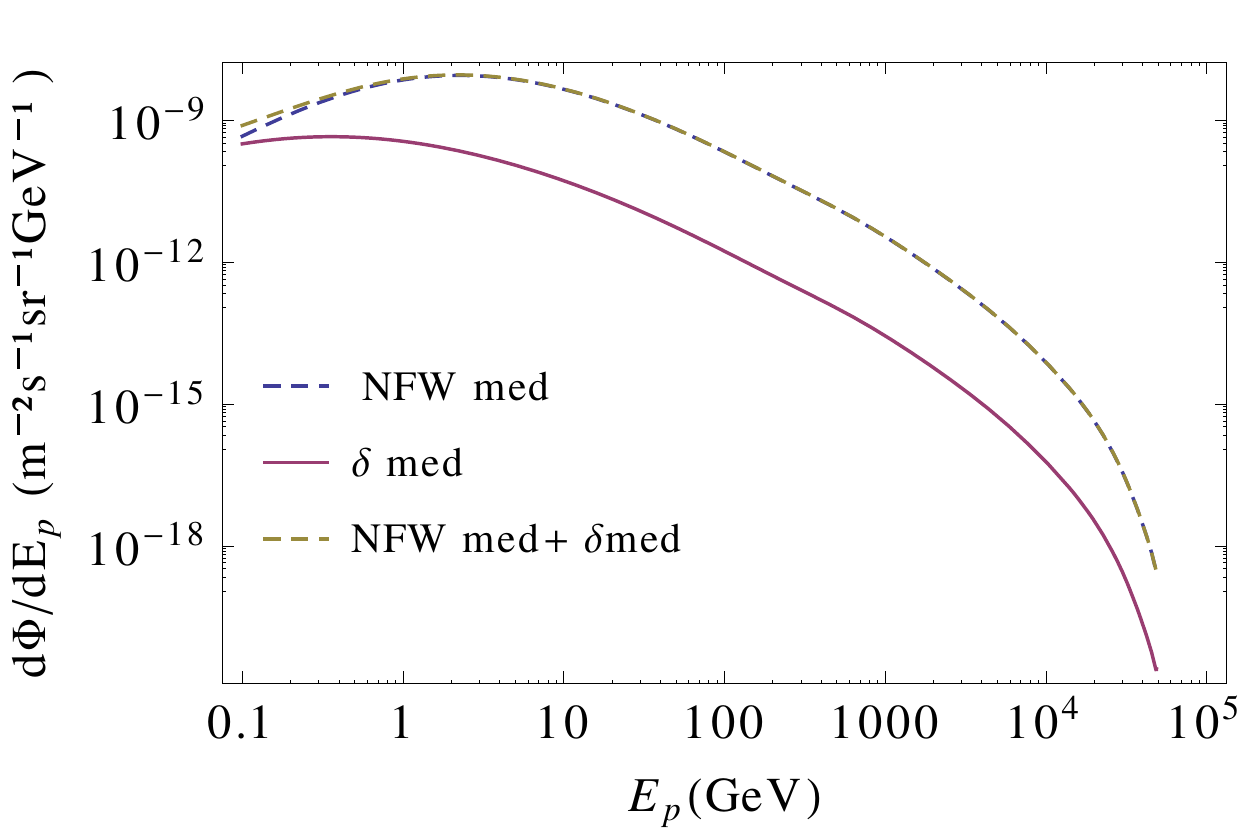}
\caption {\footnotesize{The same as Fig. \ref{NFWdeltamax} for the medium propagation model.}}
\label{NFWdeltamed}
 \end{minipage}
 \hspace{2cm}
 \begin{minipage}{.4\textwidth}
  \includegraphics[width=\textwidth,height=\textheight,keepaspectratio]{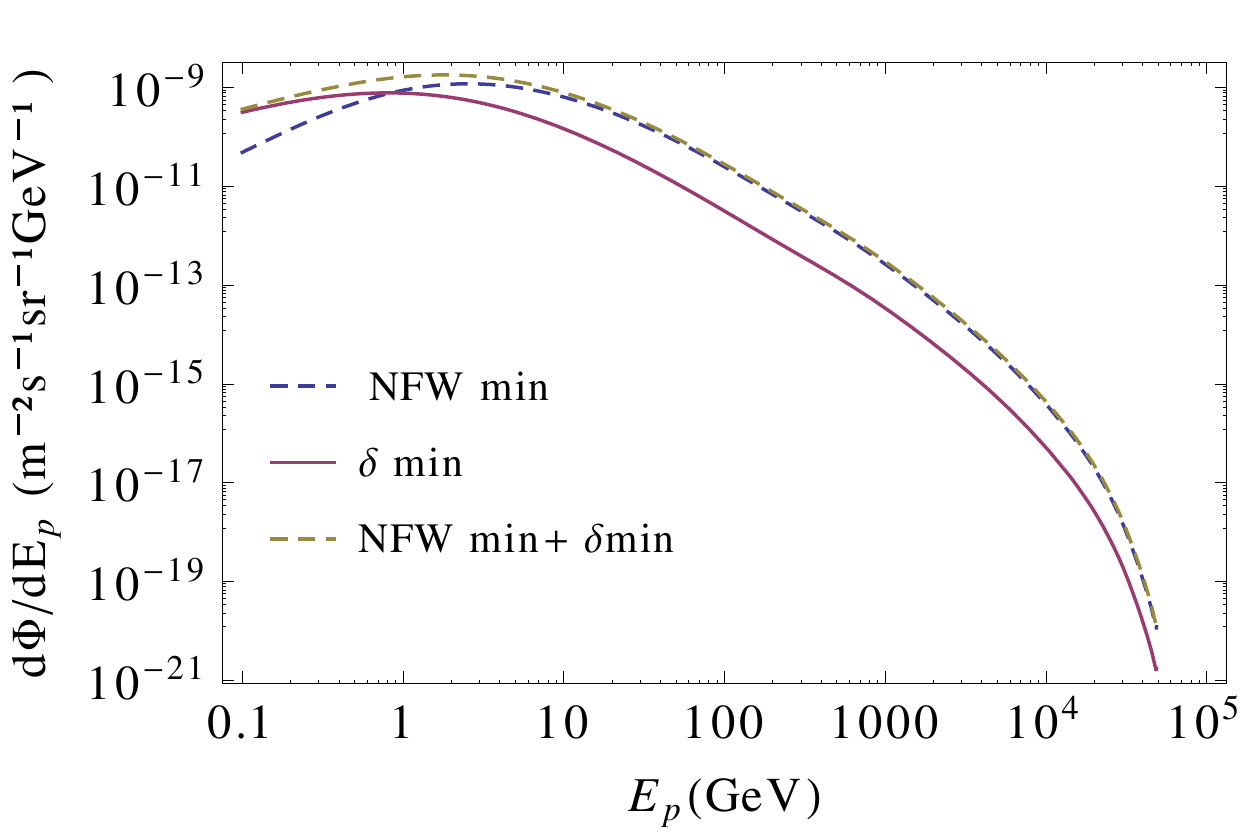}
\caption {\footnotesize{The same as Fig. \ref{NFWdeltamax} for the minimum propagation model.
In this case, the contribution from the GC is dominant at low energies.}}
\label{NFWdeltamin}
 \end{minipage}
 \end{center}
\end{figure}

The emission spectra of the source term for annihilating DM is
\begin{equation}
Q_E(\Ep)=\frac{1}{{2}}\left(\frac{\rho_\odot}{m_{\text{DM}}}\right)^2\sum_j\langle\sigma v\rangle_j\frac{dN_{\bar p}^j}{d\Ep}\,.
\end{equation}
The differential number of antiprotons per energy unit $dN_{\bar p}^j/\text{d}\Ep$,
produced in a given annihilating or decaying channel $j$, involves hadronization and possible decays of unstable products.
This requires Monte Carlo events generators \cite{Cembranos:2013cfa} or fitting or interpolation functions \cite{Ce10}. In particular, we will use
the results reported in \cite{Cirelli}. They refer to Pythia 8.135 Monte Carlo events generator software \cite{pythia} and reproduce the differential number of antiprotons  produced by DM of different masses. In this work, we will focus on antiproton fluxes coming from fragmentation and decays of SM particle-antiparticle pairs produced by DM annihilation. We will ignore DM decays. In particular, we will illustrate our analysis by
considering DM annihilation into $W^+W^-$ pairs that are consistent with the origin of the HESS J$1745$-$290$  gamma-ray observations \cite{HESSfit},
as an interesting example.

Thus, in Fig. \ref{Wpsource},
we show the antiproton flux generated by a $48.8$ TeV DM particle that annihilate into $W^+W^-$ pairs at source,  before the propagation.
As we can see, electroweak (EW) corrections are important for antiproton production at low energies.  In Fig. \ref{TOAflux}, we show the
antiproton flux at TOA after the propagation within the Galaxy when the \textit{primary} source is the mentioned $48.8$ TeV DM annihilating
into the $W^+W^-$ channel in the case of a NFW halo profile and point-like source at the GC for different diffusion models.

We can compare with the PAMELA antiproton data in order to constrain a vast range of DM masses depending on the particular value of $Q_X^0$,
the DM mass $m_{\text{DM}}$, and the annihilation or decay channel.
In particular, we will present results for the NFW halo profile and annihilating DM into the $W^+W^-$ channel by assuming the standard
thermal averaged cross-section $\langle\sigma v\rangle=3\times10^{-26}\text{cm}^{3}\text{s}^{-1}$.
In Fig. \ref{NFWDM}, we analyze the case of the only contribution of a simple NFW halo profile.

For DM masses below $m_W\simeq 80.4$ GeV, DM cannot annihilate into a real $W^+W^-$ pair. However,
for a kinematically allowed annihilation channel,
it is well known that low mass DM particles are severely constrained for masses below $1$ or $100$ GeV,
depending on the diffusion model features (minimum or maximum type, respectively).
The restrictions can be extrapolated to higher masses for high values of the boost factor $b_\text{NFW}$, which accounts
for possible enhancements of the antiproton flux due to higher annihilating cross sections or different DM density distributions.
Indeed, enhancements of order $b_\text{NFW}\simeq 10^{3}$ allow to restrict DM masses of order $m_{\text{DM}}\simeq 100$ GeV
or even $m_{\text{DM}}\simeq 10$ TeV depending on the particular features of the antiproton propagation.
It is also interesting to compare the sensitivity of the antiproton and gamma-ray
analyses. For DM particles with masses below 500 GeV, the gamma-ray
study should be done with the observations of the Fermi Gamma-ray Space Telescope. For example, we can compare
our results with the gamma-ray analysis performed in \cite{Hooper:2012sr},
although this comparison depends on the particular dark halo shape
and substructure. For a NFW halo, the thermal DM cross-section mentioned above is typically excluded for
a DM mass below 20 GeV. In the case of the $W^+W^-$ channel, this constraint does not apply since it is below
the production threshold. However, this DM model would be excluded for a small boost factor of
$b_{\text{NFW}}^{WW}=3$ at 100 GeV or $b_{\text{NFW}}^{WW}=10$ at 1 TeV. By comparing with Fig. \ref{NFWDM}, we conclude
that the gamma-ray analysis is more sensitive except for the maximum propagation model.
Under this assumption, both studies are competitive and a combined analysis could improve DM constraints.

In Fig. \ref{deltaDM}, we show the results from the same analysis computed for the propagation of antiprotons produced at a point-like
source centered at the GC. Following our convention, we need very high values of $Q_X^0$ in units of $Q_X^{0-\text{NFW}}$ in order
to find significant constraints. We can define the boost factor associated to the central contribution as
$b_{\delta-\text{NFW}}=Q_X^0/Q_X^{0-\text{NFW}}$. In contrast with $b_\text{NFW}$, it is important to remark that, in general,
there is not a preferred theoretical value for $b_{\delta-\text{NFW}}$. It depends on the particular clustering mechanism
for the DM substructure localized at the GC. Indeed, it may be much larger or smaller than one.

In such a case, we can compute the total antiproton flux as:
\begin{eqnarray}
\frac{d\Phi_{\bar p}^{\text{DM}}}{dE_{\bar p}}=
\frac{v_{\bar p}}{{8\pi}}\left(\frac{\rho_\odot}{m_{\text{DM}}}\right)^2
\sum_j\langle\sigma v\rangle_j\frac{dN_{\bar p}^j}{d E_{\bar p}}
\left[b^j_\text{NFW}\cdot R^\text{NFW}(E_{\bar p})+b^j_{\delta-\text{NFW}}\cdot R^{\delta-\text{NFW}}(E_{\bar p})\right]\,,
\end{eqnarray}
where $j$ labeled the possible different annihilating or decaying channel contribution, and $R^{\delta-\text{NFW}}(E_{\bar p})$
means the propagation function associated to the localized central contribution normalized
with $Q_X^0= Q_X^{0-\text{NFW}} \simeq 2.13\cdot10^{60}\,m^3\,\text{sr}$.
Because antiprotons sources from the GC could reach at the Earth from any direction,
they would be hardly distinguished from the ones produced by the continuous halo distribution.

\begin{figure}[t]
\begin{center}
\epsfxsize=13cm
\resizebox
{10cm}{7.5cm}
{\includegraphics{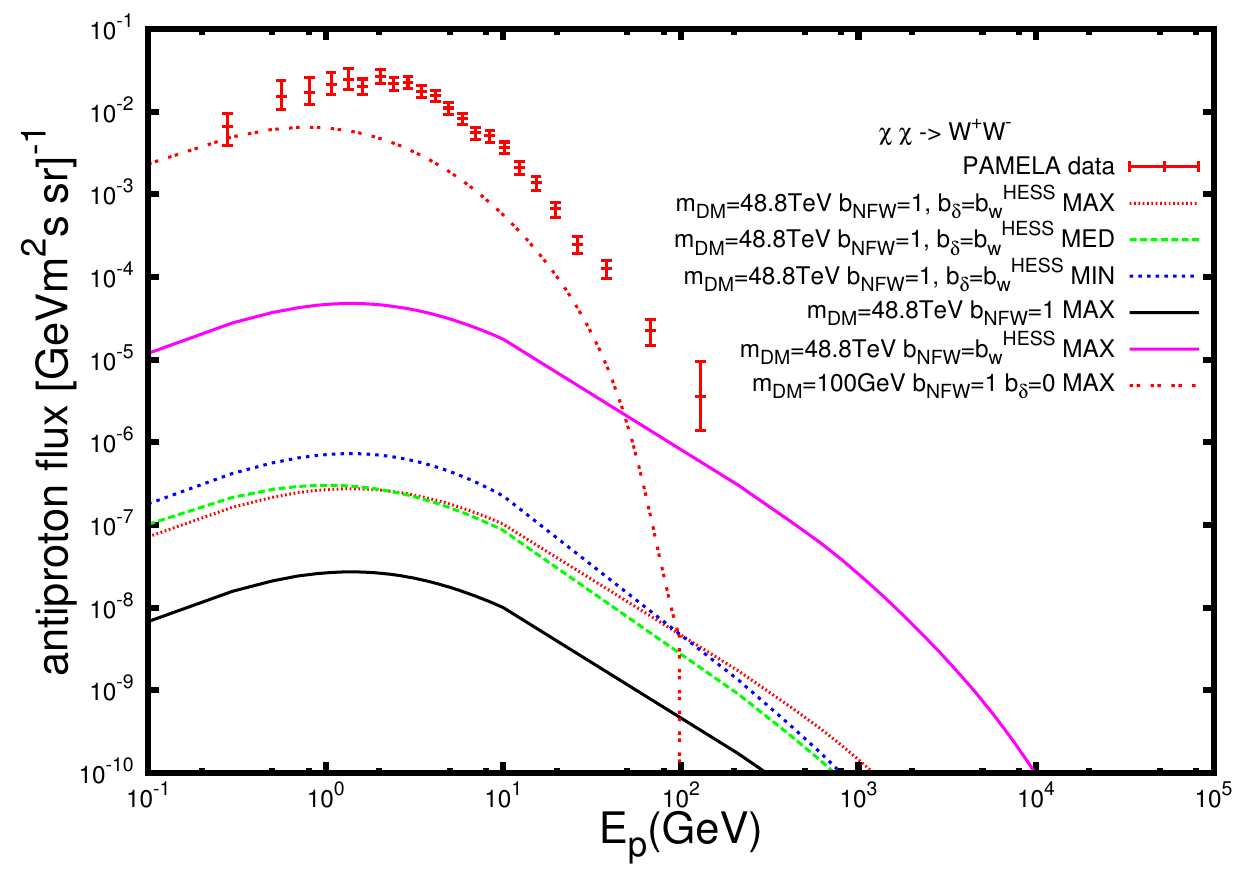}}
\caption {\footnotesize{
Antiproton differential flux at the TOA after propagation for
$48.8$ TeV DM annihilating into $W^+W^-$ pairs for different diffusion models and distribution profiles:
The lower signal (black line) corresponds to the non-boosted NFW profile by employing the maximum diffusion model.
On the other hand, the same assumptions give raise to the highest flux at high energies (violet line)
but with a boost factor of
$b^{WW}_\text{NFW}=b^{WW}_\text{HESS}= 1767$. A point-like source is negligible for this diffusion model if it is not enhanced by very large
factors
(see Fig. \ref{TOAflux} or \ref{NFWdeltamax}).
%
We show the antiproton flux at TOA for the medium diffusion model
%
for non-boosted NFW profile 
plus enhanced $\delta$-DM distribution
($b^{WW}_{\delta-\text{NFW}}=  b_{WW}^\text{{HESS}}$)
at the GC for three diffusion models (blue big-dotted, green rushed and red little-dotted line).
Finally, we show as a non-boosted NFW profile of $100$ GeV DM annihilating into the $W^+W^-$ channel is excluded
also without enhancement at the GC.
}}
\label{NFWdeltaf}
\end{center}
\end{figure}

By taking into account both contributions, and particularizing again for the $W^+W^-$ annihilation channel,
we can reach the results shown in Fig. \ref{NFWdelta} for $b^{WW}_\text{NFW} = 1$.
DM particles with masses below approximately $100$ GeV, which annihilate in the mentioned boson channel are
excluded for the maximum diffusion model. This is due to the GC contribution and
because the majority of the antiproton flux produced by heavy DM annihilating into $W^+W^-$ pairs contributes mainly at high energies.
The point-like contribution dominates for $b^{WW}_{\delta-\text{NFW}}\gtrsim 10^3$, whereas it is negligible for $b^{WW}_\delta\lesssim 10^{-2}$.
In any case, heavy DM masses are allowed also with large values of $b^{WW}_{\delta-\text{NFW}}$. In fact,
current observations are not sensitive to
the DM particle masses around $48.8$ TeV. Such a value is consistent with the origin of the HESS data of gamma-rays coming from the GC, which
requires $b^{WW}_{\delta-\text{NFW}}= 1767 \pm 419$ \cite{HESSfit}
{(We define $b^{WW}_\text{HESS}\equiv 1767$ as a benchmark boost factor, which we will use to illustrate our results).}
Similar features can be observed at Figs. \ref{NFWdeltamax}, \ref{NFWdeltamed}, \ref{NFWdeltamin}
.
The contribution from the point-like source at the GC is more important at low energies. This result is general for any value of the DM mass.
On the contrary, the mass value characterizes the emission at high energies, since the spectra show a cut-off at the DM mass. These
high energy features are quite independent of the spatial distribution assumed for the DM particles. Similar conclusions can be derived from
Fig. \ref{NFWdeltaf}, where an explicit comparison with the PAMELA data is provided.

One of the most interesting conclusions of this work is the relation between the results
associated with different analyses. In particular, by comparing Figures \ref{NFWDM} and \ref{deltaDM}, we can
observe the antiproton sensitivity dependence on the diffusion model. The most important
restrictions for the dark halo emission come from the maximum diffusion model, whereas the
minimum model is the least constraining. On the contrary, for the antiprotons produced at
the GC, the constraints are more important for the minimum diffusion model for DM masses
lower than 200 GeV. For masses higher than 1 TeV, the most constraining models are the
minimum and the maximum one.

Therefore, the most constraining diffusive model on antiproton fluxes depends
on whether they are produced from the GC (Figure \ref{deltaDM}) or from the entire halo (Figure \ref{NFWDM}).
These results can be understood by observing Fig. \ref{Rfunction} and Fig. \ref{NFWdeltaf}. The most important
constraints for the GC emission are provided by the minimum diffusion model if the
restrictions are dominated by low energy observations (antiproton data around 1 GeV). If
the restrictions come from the high energy data (around 100 GeV), minimum and maximum
propagations give the most restricting results. On the contrary, independently on the DM
mass, the maximum diffusion model is the most constraining if the antiprotons are produced
within the entire dark halo.

\section{Conclusions}

Present antiproton flux measurements are compatible with standard diffusion models of cosmic-rays without additional primary sources.
Indeed cosmic-rays interactions  with interstellar medium and their propagation represents the background for new astrophysical primaries
that may produce an important amount of antiprotons. We have analyzed the prospective signatures that should be produced by different types of antiproton spectra sources at the center of our Galaxy. The diffusion of antiproton particles highly affects the final signature.
In this sense, our analysis can be used to constrain new sources of primary antiprotons if the agreement between observations and predictions is maintained; or alternatively, it can determine the features of the diffusion model if a new antiproton flux component from the GC is identified.

We have studied the antiproton propagation function for a point-like source at the GC. In general, this function depends on the spatial distribution
source. We have analyzed the flux at the TOA for three emission spectra as different models that could be associated to a large variety of astrophysical sources, such as the case of a power law flux, monochromatic emission, or to annihilating or decaying DM.
We have compared such flux with the present antiproton data. In order to be conservative, we have neglected the background contribution.
We have studied the sensitivity by constraining the different features of the mentioned spectra, as the total normalization amplitude, the power index, the characteristic energy, the DM mass, etc. The constraints are very general and need to be compared with particular motivated sources. Alternatively, if an excess is observed, our analysis can determine the particular model favoured for such data.

In the case of the DM, there are two reasons for the analysis. On the one hand, DM can be compacted around a
very localized region around the center of our galaxy for different processes, as the baryonic compression or
black hole effects. On the other hand, numerical computations of the diffusion equation present a singularity
at the GC, because of the central behaviour of DM halos density profiles. This divergence needs to be regularized.
The simplest possibility is removing it below a given radius \cite{Cirelli}. Another possibility is to consider
its contribution separately as a point-like source. In any case, the contribution from the local
continuous halo profile is expected to be important and the interplay between both contributions gives a rich phenomenology,
as we have shown in the present study.

\vspace{0.5cm}

\acknowledgments

This work has been supported by UCM predoctoral grant, MICINN (Spain) project numbers FIS 2008-01323, FIS2011-23000, FPA2011-27853-01,
Consolider-Ingenio MULTIDARK CSD2009-00064, and the Department of Energy, Contract DE-AC02-76SF00515.



\end{document}